\documentclass[reprint,aps,pra,showpacs,superscriptaddress]{revtex4-1}
\usepackage{amssymb,amsmath}
\usepackage{graphicx}
\usepackage{dcolumn}
\usepackage{multirow}
\usepackage{color}

\definecolor{purple}{rgb}{0.61,0.19,1.00}

\begin{document}
	\title{Compressed Optimization of
		Device Architectures (CODA) for semiconductor quantum devices}
	\author{Adam Frees}
	\email[Correspondence should be addressed to: ]{frees@wisc.edu}
	\affiliation{Department of Physics, University of Wisconsin-Madison, Madison, WI 53706}
	\author{John King Gamble}
	\affiliation{Center for Computing Research, Sandia National Laboratories, Albuquerque, NM 87123}
	\affiliation{Quantum Architectures and Computation Group, Microsoft Research, Redmond, WA 98052}
	\author{Daniel R. Ward}
	\affiliation{Center for Computing Research, Sandia National Laboratories, Albuquerque, NM 87123}
	\author{Robin Blume-Kohout}
	\affiliation{Center for Computing Research, Sandia National Laboratories, Albuquerque, NM 87123}
	\author{M. A. Eriksson}
	\author{Mark Friesen}
	\author{S. N. Coppersmith}
	\affiliation{Department of Physics, University of Wisconsin-Madison, Madison, WI 53706}

	\begin{abstract}
		Recent advances in nanotechnology have enabled researchers to manipulate small collections of quantum mechanical objects with unprecedented accuracy.
		In semiconductor quantum dot qubits, this manipulation requires controlling the dot orbital energies, tunnel couplings, and the electron occupations. These properties all depend on the voltages placed on the metallic electrodes that define the device, whose positions are fixed once the device is fabricated.
		While there has been much success with small numbers of dots, as the number of dots grows, it will be increasingly useful to control these systems with as few electrode voltage {changes} as possible.
		Here, we introduce a protocol, which we call the Compressed Optimization of Device Architectures (CODA), in order to both {efficiently} identify sparse sets of voltage changes that control quantum systems, and to introduce a metric which can be used to compare device designs.
		As an example of the former, we apply this method to {simulated devices with up to 100 quantum dots and show that CODA automatically tunes devices more efficiently than other common nonlinear optimizers.}
		{ }To demonstrate the latter, we determine the optimal lateral scale for a triple quantum dot, yielding a simulated device that can be tuned with small voltage changes on a limited number of electrodes.
	\end{abstract}
	
	\maketitle

	\section{Introduction}
	
	Nanoscale systems are challenging to control in part because their size makes them susceptible to even the smallest materials defects.
	Quantum devices present special challenges because their energy spectra and tunnel couplings {each} require precise control \cite{Hanson:2007p1217,Zwanenburg:2013p961}.
	Here{,} we focus on quantum bits (qubits) formed in
	electrostatically-gated quantum dots \cite{Loss:1998p120}. 
	In these systems, voltages are simultaneously tuned on many electrodes to precisely shape the electrostatic potential landscape within a device.
	Working with a small number of qubits, researchers have already demonstrated excellent qubit
	coherence and performance in devices based in GaAs \cite{PhysRevLett.105.246804,PhysRevLett.110.146804,PhysRevLett.116.086801,PhysRevLett.116.116801} and silicon \cite{Kim:2014p70,KimNatNano15,PhysRevLett.116.110402,Kawakami11738,Schoenfield:2017aa,Thorgrimsson:2017aa,Yoneda:2018aa,Samkharadze1123,Mi:2018aa,PhysRevLett.120.137702,Jock:2018aa},
	including the successful implementation of
	two-qubit gates \cite{Veldhorst:2015aa,Nichol:2017aa,Zajac439}
	and algorithms \cite{Watson:2018aa}.
	Additionally, there has been rapid progress in systems with electrons bound to donors \cite{Pla:2012p541,Fuechsle:2012p242,Pla:2013p334,doi:10.1063/1.4930909,PhysRevLett.115.166806,Laucht:2016aa,Watsone1602811,Tosi:2017aa,Harvey-Collard:2017aa,Broome:2018aa}, which share many of the same design challenges as quantum dot qubits.
	Tuning schemes for these devices are typically determined empirically; however,
	there has been recent progress towards automatic tuning of quantum dots both experimentally \cite{BaartAPL,doi:10.1063/1.5031034} and in simulated systems \cite{10.1371/journal.pone.0205844}.
	One challenge in controlling quantum dot devices is ensuring that the voltage changes on the electrodes remain small during tuning, a property that we refer to as \emph{voltage moderation}. Imposing voltage moderation reduces the power required during dynamic operation and lowers the risk of instability. Another challenge is ensuring that a small number of electrodes can be used to tune quantum dots{, preferably with those electrodes proximal to the relevant dots}. We refer to this property as \emph{voltage sparsity}, which eases the demands on control electronics and will be increasingly important as devices are scaled to very large numbers of quantum dots.
	In order to achieve these goals, it {is critical} to use simulations both to identify moderate and sparse tunings, and {to design these features into devices before they are fabricated.}
	
	Here, we introduce the Compressed Optimization of Device Architectures (CODA) protocol, which both determines {optimized ways} to change the voltages in a given system to achieve a desired outcome, and provides a metric to characterize the ease with which the device can be tuned.
	We show that by minimizing the $L_1$ norm of the applied voltage changes, we can simultaneously achieve voltage moderation and voltage sparsity.
	We minimize this norm by implementing the CODA protocol, which relies on results and methods used for compressed sensing \cite{Candes:2006p1207,Donoho:2006p1289} in the signal processing literature.
	Using a simulated eight-dot device, we demonstrate that CODA yields solutions that are simultaneously sparse and moderate. Moreover, we show that by imposing voltage sparsity and moderation, we obtain solutions that only involve gates that are proximal to the dot being manipulated -- an extremely desirable property for extensibility. {To further demonstrate the extensibility of the CODA protocol, we use a simple model to measure the number of device simulations required to tune devices with up to 100 quantum dots. We find that CODA requires nearly an order of magnitude fewer simulations than other commonly-used nonlinear optimization techniques.}
	{Additionally}, formulating control as an optimization problem allows us to directly compare the effectiveness of different device architectures, enabling improvement of the electrode designs themselves. To demonstrate this, we use CODA to optimize the overall lateral scale of a triple quantum dot, which leads to a device that is optimally ``tunable."
	
	\section{Automatic tuning of simulated devices}

	\begin{figure*}[tb]
		\includegraphics[width= 0.95 \linewidth]{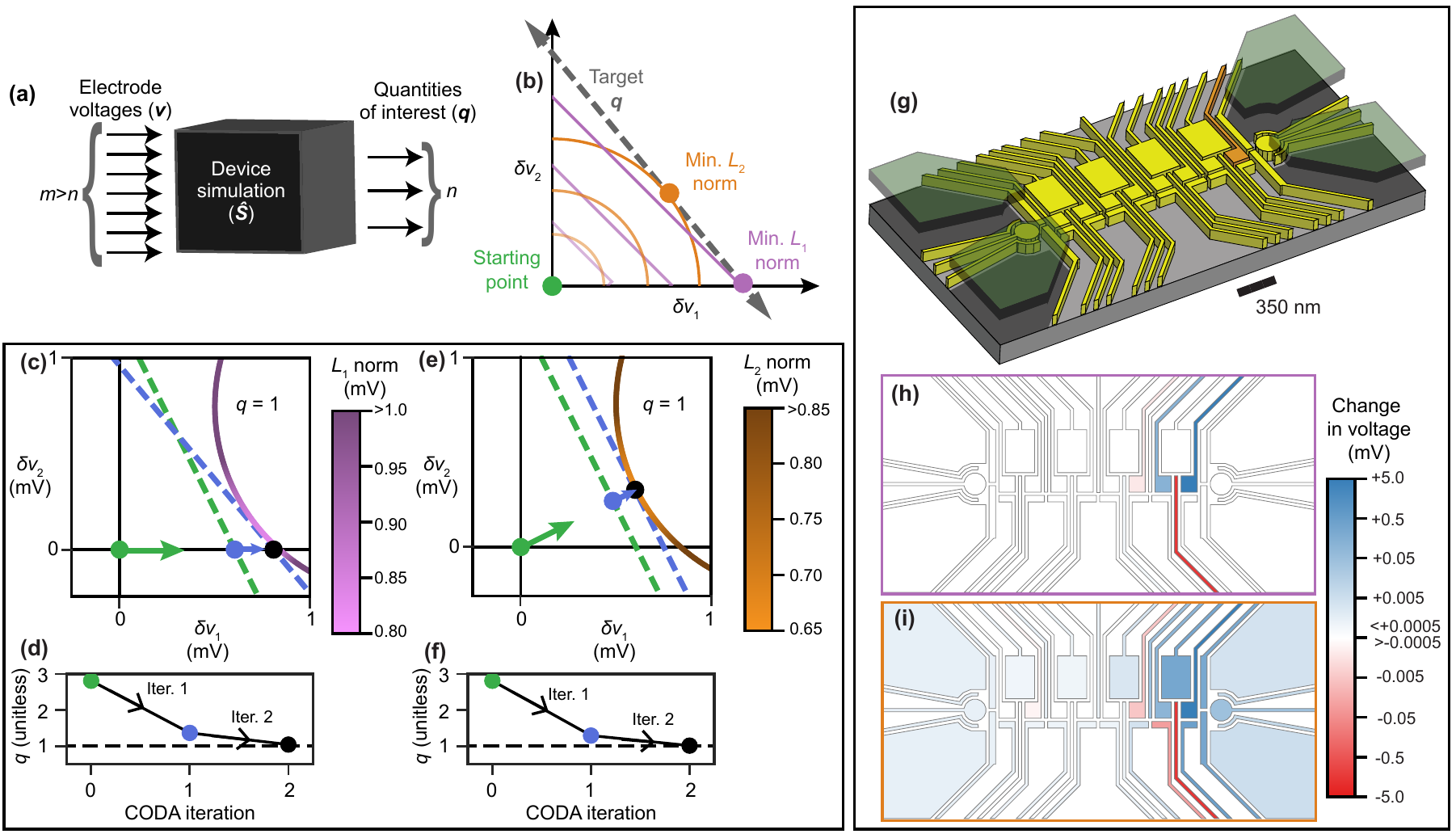}
		\caption{\label{fig1} 
			Example implementations of the CODA protocol, including comparisons of using the $L_1$ norm (the sum of absolute values) and the $L_2$ norm (the Euclidian length) of the voltage changes needed to achieve the desired values of the quantities of interest (dot occupations, tunnel couplings, \emph{etc.}).
			(a) A simulated quantum dot device maps electrode voltages to quantities of interest. Generically, there are more electrodes in a device than quantities of interest.
			(b) There are many combinations of electrode voltages that result in a target system state (grey dashed line). By choosing the solution associated with the minimum $L_1$ norm of voltage changes, described in Eq.~\eqref{eq1} (purple), we simultaneously ensure that voltages are changed by small amounts on a small number of electrodes. Minimizing the $L_2$ norm (orange) does not ensure that voltage changes will be applied to a small number of electrodes.
			(c)-(f) Depiction of CODA algorithm tuning voltages to obtain $q_\text{target} =1$ in a simple system. 
			The voltages which yield the target quantity of interest are indicated with the circular segment in panels (c) and (e).
			Starting at $\delta v_1=\delta v_2= 0$ mV ({green} circle), we calculate the Jacobian $\mathbf{J}$ and find all of the solutions to the linear equation $q_\text{target} - q_\text{current} = \mathbf{J}\cdot\delta \mathbf{v}$, shown as a {green} dashed line.
			We then find the voltage changes on this line which minimize the $L_1$ and $L_2$ norm, {blue} circle in panel (c) and (e), respectively. Using the derivatives at this new point, we again estimate the voltage changes required to hit the target ({blue} dashed line), and again find the solution which minimizes the appropriate norm (black circle). 
			The solution found here has converged on $q_\text{target}$.
			Additionally, we have obtained the voltage changes associated with the minimum of the appropriate norm, indicated by the purple color scale in panel (c), and the orange color scale in panel (e).
			(g)-(i) CODA tuning of an 8-dot device.
			(g) Schematic of a simulated 8-dot device, with metal electrodes colored yellow/orange (lower level) and green (upper level). Here, the objective is to increase the occupation of the right-most quantum dot (underneath the orange electrode) by one electron, keeping all other dot occupations and tunnel couplings unchanged.
			(h),(i) Visualization of the voltage changes obtained by the optimization protocol, plotted on a logarithmic color scale (electrodes with voltage changes less than 0.5 $\mu$V are colored white), minimizing the (h) $L_1$ and (i) $L_2$ norms of the voltage changes. As expected, minimizing the $L_1$ norm ensures that a limited number of electrode voltages are changed, whereas minimizing the $L_2$ norm does not.
		}
	\end{figure*}
	
	For qubit applications, the main properties we wish to control are the quantum dot occupations, the orbital energy splittings, and the tunnel rates between the dots. 
	Such properties are referred to as \emph{quantities of interest}, which we represent as
	a vector $\mathbf q$ in a vector space $\mathcal Q$.
	These values are controlled by the voltages applied to the electrodes.
	A given set of voltages is represented as a vector $\mathbf v$ in a vector space $\mathcal V$.
	A physical system, or a simulation thereof, acts as a function that maps the voltages to the quantities of interest: $\hat S: \mathcal V \rightarrow \mathcal Q$,
	as shown in Fig.~\ref{fig1}(a). 
	
	Because generically there are more electrodes than quantities of interest, many different $\mathbf v$ can yield target values for the quantities of interest $\mathbf q_\text{target}$.
	The solutions are not all equivalently useful -- it is our goal to select changes in voltages that are simultaneously moderate and sparse.
	In Fig.~\ref{fig1}(b), we highlight the advantage of choosing the solution that satisfies the equation
	\begin{equation}
	\widetilde{\delta \mathbf v} = \underset{\delta \mathbf v} {\text{argmin}} \left|\left| \delta \mathbf v \right|\right|_1,
	\mbox{ subject to }\hat S(\mathbf v_\text{init} + \delta \mathbf v) = \mathbf q_\text{target},\label{eq1}
	\end{equation}
	where $\mathbf v_\text{init}$ is the voltage vector at the starting point, $\mathbf q_\text{target}$ is the vector of target quantities of interest,
	$\delta \mathbf v$ is the voltage change from the starting point, and 
	$|| \cdot ||_1$ is the $L_1$ norm, which is the sum of the absolute value of each element in the vector.
	
	Minimizing the $L_1$ norm of the voltage change vector ensures both the magnitude of the individual voltage changes remain small (\emph{i.e.}, voltage moderation) and that the voltage change vector is sparse. The sparsity achieved by $L_1$ norm minimization is a property used extensively in the field of compressed sensing \cite{Candes:2006p1207,Donoho:2006p1289}. 
	In contrast, $L_2$ norm minimization (minimizing the Euclidian length of $\delta \mathbf v$) does not guarantee voltage sparsity, and $L_0$ pseudonorm minimization (minimizing the number of nonzero elements of $\delta \mathbf v$) does not guarantee voltage moderation. {In principle, one could achieve both moderation and sparsity by minimizing some combination of these two quantities, but this would involve making an arbitrary choice for the relative weight given to the $L_0$ and $L_2$ norms.} Additionally, finding the solution that minimizes the $L_0$ pseudonorm is known to be an NP-hard problem \cite{Natarajan:1995p227}, whereas convex programming methods can be used to minimize the $L_1$ norm efficiently \cite{JMLR:v17:15-408}.
	Therefore, by selecting the changes in voltages described in Eq.~\eqref{eq1}, the CODA procedure yields a device tuning in which a small number of voltages are changed by modest amounts.
	
	To demonstrate the CODA protocol, we first perform automatic tuning of a simple {toy} example shown in Fig.~\ref{fig1}(c)-(f). This system comprise{s} two electrode voltages $\delta v_1$ and $\delta v_2$ and one quantity of interest $q(\delta v_1,\delta v_2) = (\delta v_1/(1 \text{ mV})-3/2)^2+(\delta v_2/(1 \text{ mV})-3/4)^2$. During the protocol, we treat this as a black box function, as in more complicated device models we do not have access to an analytical form of the mapping from voltages to quantities of interest. Starting with $\delta v_1=\delta v_2= 0$ mV, we wish to find the voltage changes with minimal $L_1$ norm that are necessary to change the quantity of interest from $q_\text{current} = 2.8125$ to $q_\text{target} = 1$.
	To achieve this, we apply an iterative algorithm to minimize the distance between the simulated quantities of interest and the target quantities of interest.
	At the starting point, we calculate the Jacobian $\mathbf{J} = (\frac{\partial q}{\partial \delta v_1},\frac{\partial q}{\partial \delta v_2})$ and find all of the solutions to the linear equation $q_\text{target}-q_\text{current} = \mathbf{J}\cdot\delta \mathbf{v}$, shown as a red dashed line in Fig.~\ref{fig1}(c). In this example, it is easy to minimize the $L_1$ norm along this line; however, for more general and complicated problems, we employ a matrix-free conic optimization algorithm \cite{JMLR:v17:15-408} to determine the particular $\delta \mathbf{v}$ that minimizes the $L_1$ norm while satisfying this equation. This vector defines a search direction in voltage space; CODA moves along this direction in voltage space, stopping at the point that brings the simulation closest to the target quantities of interest. This process then repeats until the quantity of interest converges on $q_\text{target}$. {In general, convergence is declared when $\|\mathbf{q}-\mathbf{q}_\text{target}\|_2$ falls below some threshold. We typically choose an error threshold of $\sim 10^{-2}$; in this example, we choose a threshold of $5 \times 10^{-2}$. We achieve} convergence after two iterations, at which point $q \approx 1.041$, as shown in Fig.~\ref{fig1}(d). In this example, we achieved the target quantity of interest by changing only one electrode voltage. 
	
	It is instructive to contrast this solution to one obtained using an analogous procedure in which the $L_2$ norm is minimized, rather than the $L_1$ norm, as demonstrated in Fig.~\ref{fig1}(e),(f).
	Here, we again declare convergence after two iterations, with $q\approx1.012$. However the final solution involves changing the voltage on both electrodes to achieve the target quantity of interest, and is therefore less sparse than the solution obtained using the $L_1$ norm. For more details {on the CODA algorithm}, see Appendix A.
	
	In a more realistic demonstration of CODA's capabilities, we can use it to perform automatic tuning of the accumulation-mode eight-dot device shown in Fig~\ref{fig1}(g).
	The device contains four capacitively-coupled double quantum dot qubits in a Si/SiGe heterostructure.
	{In principle, the CODA protocol can use any underlying physical model.}
	Here, we model the device using the semiclassical Thomas-Fermi approximation \cite{Stopa:1996p13767} to compute electron densities and potentials, and the WKB approximation \cite{shankar1994principles} to calculate tunnel rates, as described in Appendices B-C. The quantities of interest are chosen to be the eight dot occupations and the four intra-qubit tunnel rates.
	{While this semiclassical approach is appropriate for these quantities of interest, one may need a more sophisticated model to correctly capture the physics of different quantities of interest, \emph{e.g.} valley splitting within a dot \cite{PhysRevB.70.165325}.}
	Our starting point is chosen to give dot occupations of 1 electron, and transmission coefficients between the dots of 0.01, corresponding to tunnel rates of approximately 400 MHz
	(see Appendix B-C for details about the simulation parameters and methods). Our goal is to find the optimal changes in voltages that can be applied to the device electrodes to add one electron to the right-most dot, leaving all other dot occupations and transmission coefficients unchanged.
	
	In Fig.~\ref{fig1}(h), we show the voltage changes needed to achieve convergence of the CODA procedure  applied to the eight-dot device. Note that although voltages are allowed to vary on all the electrodes, CODA chooses to vary only four electrode voltages, and those electrodes are proximal to the target of interest.
	For comparison, we also performed an alternative optimization protocol based on $L_2$-minimization, with results shown in Fig.~\ref{fig1}(i). Note that although the latter protocol achieves the same target quantities of interest, the solution involves voltage changes on almost all of the electrodes, indicating that this solution is neither sparse nor proximal.
	In both tuning protocols, the magnitudes of the voltage changes applied to the electrodes are all under 5 mV. Because minimizing the $L_2$ norm explicitly ensures voltage moderation, the similarity in the magnitude of voltage changes applied in both cases confirms that the solution found via $L_1$ norm minimization exhibits voltage moderation as well.
	Hence, we confirm that CODA is a practical tool 
	for tuning a device,
	because it selects voltages that are both sparse and moderate while achieving the specified target quantities of interest.
	
	{\section{Extensibility}}
	{In the previous section, we applied the CODA protocol to simulated devices with up to eight quantum dots, showing that the method can find sparse voltage tunings which are moderate and sparse for these devices. These solutions were obtained after two iterative steps, suggesting that this procedure is efficient even for large systems.
		Here, we show that the CODA protocol scales efficiently with device size, and can be used to find sparse tunings for devices with 100 quantum dots. }
	\begin{figure}[tb]
		\includegraphics[width=  \linewidth]{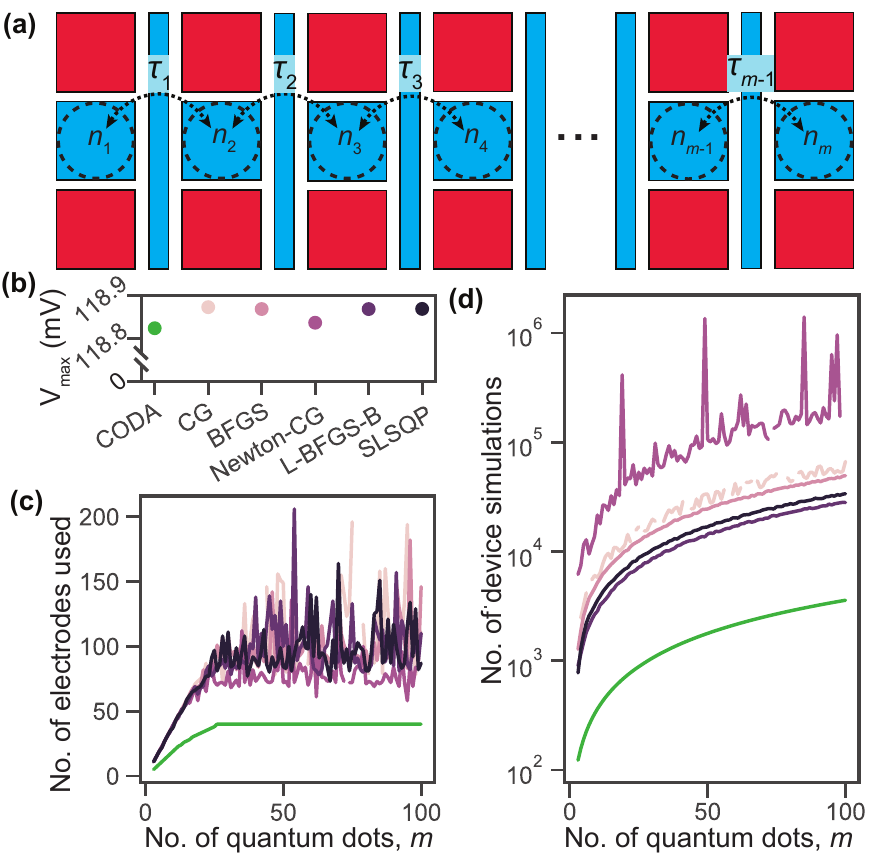}
		\caption{\label{fig4} {
				Demonstrating the extensibility of the CODA protocol using a simple model involving  up to 100 quantum dots.
				(a) Diagram of simulated device. The quantities of interest in this device are the occupations $n_i$ of the quantum dots (dashed circles) and tunnel rate $\tau_i$ between the dots (dashed double arrows). Each quantum dot has three electrodes in close proximity: one located directly above the quantum dot (blue square) and two located above and to either side of the quantum dot (red squares). Additionally, there is an electrode separating each pair of dots (thin blue rectangle). The dependence of $n_i$ and $\tau_i$ on the electrode voltages is defined phenomenologically in Eqs.~\eqref{simple_occ} and \eqref{simple_T}, respectively. 
				(b)-(d) Given a device with $m$ quantum dots, we use a variety of nonlinear optimizers (including CODA) to find the changes in electrode voltages which add one electron to the left-most quantum dot, keeping all other occupations and tunnel rates constant. We consider devices with $m$ ranging from 2 to 100. The results corresponding to CODA are shown in green, and the results corresponding with the other nonlinear optimizers are shown in hues of purple. In panel (b) we show the average over all considered devices of the maximum voltage change applied to any of the electrodes. The standard deviation is smaller than the points used in this plot. The uniformity of the results here indicates that all of the optimizers find solutions with similar voltage moderation.
				In panel (c) we show the number of electrodes used in each solution. The CODA procedure consistently requires fewer electrodes than any of the other nonlinear optimizers considered.
				In panel (d) we show the number of function calls used by each optimizer. CODA requires roughly an order of magnitude fewer function calls than any of the other nonlinear optimizers considered.}
		}
	\end{figure}

	{
		We consider the device shown in Fig.~\ref{fig4}(a), which consists of $m$ quantum dots, $m-1$ tunnel rates, and $4m-1$ electrodes. The electrode separating the quantum dots (thin blue rectangle in Fig.~\ref{fig4}(a)) is 35 nm from the center of the neighboring electrodes (red and blue squares in Fig.~\ref{fig4}(a)). The centers of the square-shaped electrodes are separated by 50 nm. The quantum dots are located 20 nm below the electrodes.}
		
		{There are many methods one could employ to model this device, including taking the semiclassical approach described in Appendix B-C, or self-consistently solving the Schr{\"o}dinger and Poisson equations, which more accurately take into account quantum effects. Here, we use a simple model that can be regarded as phenomenological, although it is physically motivated, describing a non-linear system in which an electrode's proximity to a quantum dot or tunnel barrier determines that electrode's effect on the corresponding quantity of interest.}
	{Specifically, we define the occupation $n_i$ of dot $i$ to be
		\begin{equation}
		n_i = \sum_j \frac{\left(V_j/(1\text{ mV}) + \frac{1}{10}\text{sgn}(V_j){V_j}^2/(1\text{ mV}^2)\right)}{\left(\|\overrightarrow{r_{V_j}}-\overrightarrow{r_{n_i}}\|_2\right)^3/(1\text{ nm}^3)},\label{simple_occ}
		\end{equation}
		and the tunnel rate $\tau_i$ between the $i^\text{th}$ and $(i+1)^\text{th}$ quantum dot is given by
		\begin{equation}
		\tau_i = \frac{1}{100}\sum_j \frac{\left(V_j/(1\text{ mV}) + \frac{1}{2}\text{sgn}(V_j){V_j}^2/(1\text{ mV}^2)\right)}{\left(\|\overrightarrow{r_{V_j}}-\overrightarrow{r_{n_i}}\|_2\right)^3/(1\text{ nm}^3)}. \label{simple_T}
		\end{equation}
	In these equations, $V_j$ is the voltage applied to electrode $j$, $\text{sgn}(x)$ is the sign function of $x$, $\overrightarrow{r_{V_j}}$ is the center position of electrode $j$, $\overrightarrow{r_{n_i}}$ is the center position of dot $i$, and $\overrightarrow{r_{\tau_i}}$ describes the half-way point between dots $i$ and $(i+1)$. The $r^{-3}$ scaling of these quantities is the expected spatial decay for a screened 2DEG \cite{davies_1997}, while the voltage dependence was chosen assuming that these quantities scale approximately linearly with voltage, with an additional, explicitly non-linear contribution. }
	
	{To study the extensibility of our approach, we employ a variety of nonlinear optimizers, including CODA, to tune the voltages in devices with $m=2$ to 100 dots. In each case, we begin with voltages -100 mV applied to each of the electrodes indicated with red in Fig.~\ref{fig4}(a), and with positive voltages applied to the electrodes indicated with blue in Fig.~\ref{fig4}(a). The latter voltages are set such that $n_i = 1$ and $\tau_i = 0.01$ for every $i$. We then find a combination of voltage changes that adds one electron to dot $i=1$, keeping all other $n_i$ and $\tau_i$ fixed. We specifically consider the CODA protocol, the conjugate gradient (CG) algorithm, the Broyden, Fletcher, Goldfarb, and Shanno (BFGS) algorithm, the Newton conjugate gradient (Newton-CG) algorithm, the limited memory BFGS algorithm L-BFGS-B, and the Sequential Least SQuares Programming (SLSQP) algorithm, as implemented in the SciPy package \cite{Jones:2001aa}. In all of these algorithms we minimize $\|\mathbf q - \mathbf q_\text{target}\|_2$, where $\|\cdot \|_2$ denotes the $L_2$ norm, $\mathbf q$ is the vector consisting of every quantity of interest, $n_i$ and $\tau_i$, and $ \mathbf q_\text{target}$ is the vector consisting of the target values for the quantities of interest. We define the system to be converged on the target when $\|\mathbf q - \mathbf q_\text{target}\|_2<10^{-5}$.}
	
	{We assess the voltage moderation and sparsity of each of the solutions found by the nonlinear optimizers. The maximum voltage change $V_{max}$ applied to a given device is shown in Fig.~\ref{fig4}(b). The number of nonzero voltage changes found by each optimizer as a function of $m$ is shown in Fig.~\ref{fig4}(c). While all of the nonlinear solvers apply voltage changes of comparable magnitudes, CODA consistently finds solutions which require changing fewer electrode voltages than any of the other optimizers we consider. Moreover, the number of electrodes used by CODA does not change for devices with $m\geq 26$. In contrast, all of the other nonlinear optimizers show high variability in the number of electrodes used as a function of device size. }
	
	{We find that the CODA protocol requires fewer function calls than any of the other nonlinear optimizers considered. In Fig.~\ref{fig4}(d), we show the number of function calls made in each optimization.
		 The average number of function calls made by a given nonlinear optimizer scales linearly with the number of quantum dots in the system, regardless of the optimizer used. In the case of the CODA protocol (green line in Fig.~\ref{fig4}(d)), the number of function calls is exactly proportional to the device size, as the number of iterations required for CODA to converge is independent of the number of quantum dots.
		 Although the overall scaling is the same for each of these optimizers, the CODA protocol achieves convergence using approximately an order of magnitude fewer function calls compared with the next-most efficient optimizer, the L-BFGS-B algorithm. Since the device simulations dominate the computation time, the CODA protocol can therefore automatically tune simulated devices significantly faster than any of the other nonlinear optimizers considered. We conclude that CODA can efficiently find sparse and moderate solutions in large devices.}
	
	\vspace{0.5cm}
	\section{Device design optimization}
	
	\begin{figure}[tb]
		\includegraphics[width= 0.95 \linewidth]{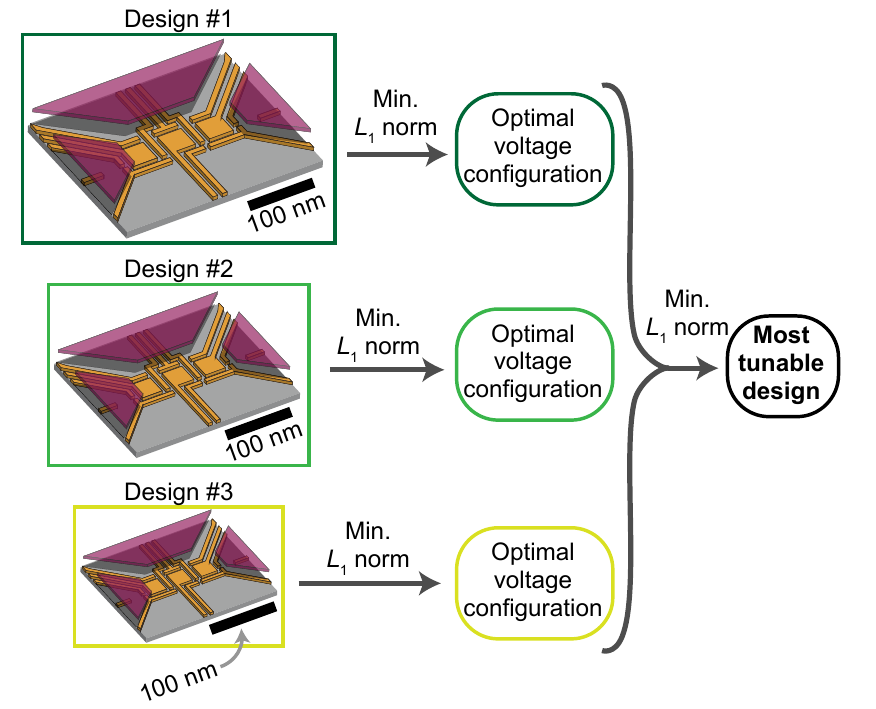}
		\caption{\label{fig2} 
			The protocol used to compare the ``tunability" of device designs. Given multiple simulated devices, we use the CODA protocol to find the minimum $L_1$ norm of voltage variations needed to induce a common change in each device (\emph{e.g.}, change the dot occupations in device \#1, device \#2, \emph{etc.}). The device with the minimum norm can simultaneously provide voltage moderation and sparsity, and should therefore be regarded as the most ``tunable" device. 
		}
	\end{figure}
	
	\begin{figure}[tb]
		\includegraphics[width= 0.95 \linewidth]{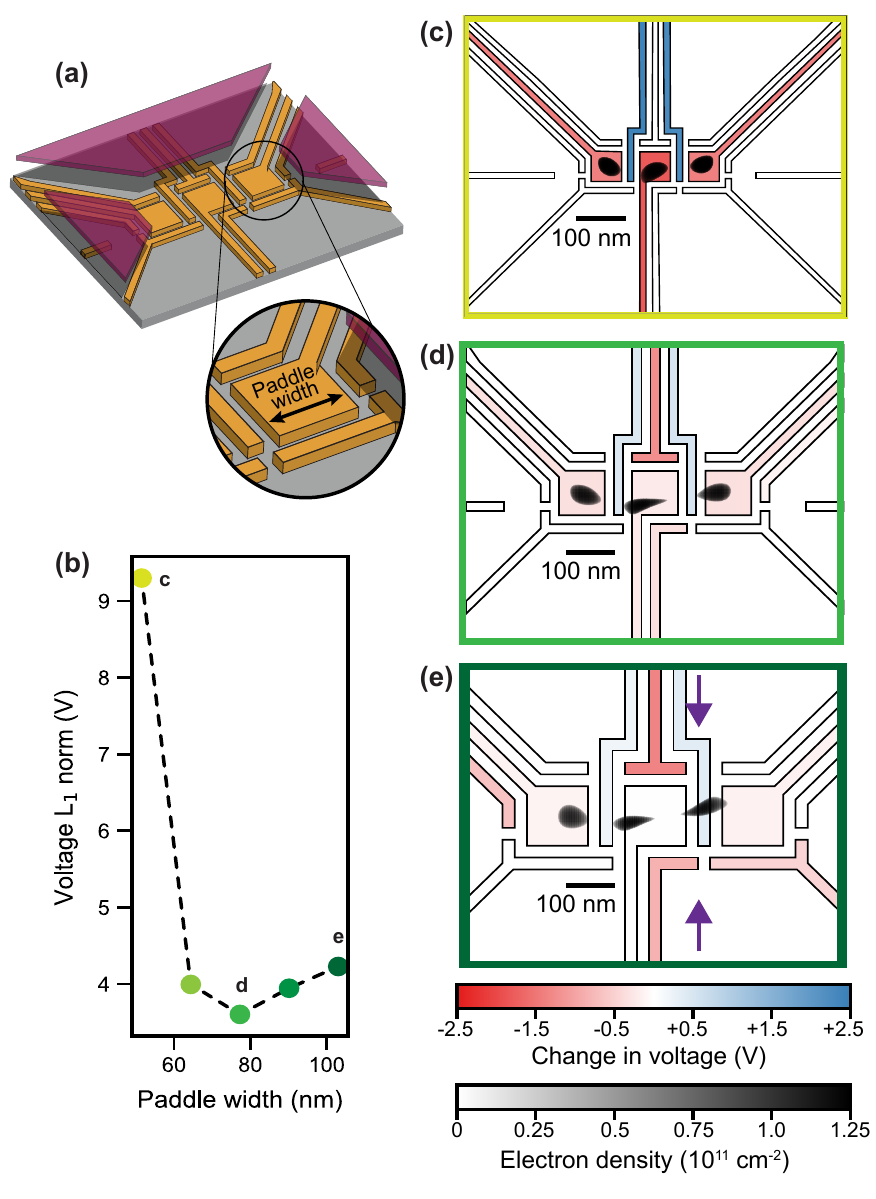}
		\caption{\label{fig3} 
			Using CODA to optimize the triple quantum dot designs shown in Fig.~\ref{fig2}.
			(a) Every device in the series has an identical electrode layout, save for the lateral scale of the device, which we characterize in terms of the width of the paddle electrode. The Si/SiGe heterostructures for all devices considered are identical.
			(b) For each device in the series, we use the CODA protocol to lower all three dots from an occupation of thirty electrons to one electron, keeping the transmission coefficients fixed at 0.01. Additionally, the single-electron dots are required to have orbital excitation energies greater than 1 meV. 
			(c)-(e) Visualization of the optimized voltage variations required to tune each device. Voltage changes are shown in red and blue, and the resulting electron density distributions are shown in black.
			For smaller devices (\emph{e.g.}, (c)), fewer electrodes are required to tune the device. However larger voltage changes must be applied to those electrodes, resulting in a high voltage $L_1$ norm. For larger devices (\emph{e.g.}, (e)), quantum dots no longer form underneath the paddle electrodes (\emph{e.g.}, right-most dot, indicated with purple arrows), so that many electrodes are required to tune the device. Balancing these effects leads to a local minimum in the voltage $L_1$ norm, corresponding to device (d).
		}
	\end{figure}
	
	In addition to automatically tuning quantum dot devices, 
	the CODA protocol can be used to characterize the voltage sparsity and moderation of typical gate operations, thus providing a key metric for evaluating and comparing different device designs.
	Here, we consider a series of triple-dot devices, shown in Fig~\ref{fig2}. Each device has the same electrode layout, save for the overall lateral scale -- we parameterize this scale via the width of the paddle electrode, as shown in Fig~\ref{fig3}(a). All devices have an identical Si/SiGe heterostructure with a silicon quantum well a distance 30 nm below the electrodes. Optimizing device size is important, because a device with electrodes too small will lead to {instability} and larger power requirements for switching, and a device with electrodes too big will not have sufficient control over the potential landscape at the required length-scales. In particular, it has been observed in experiments \cite{Veldhorst:2015aa,Ward:2016aa,Watson:2018aa} that in larger devices it is often necessary to form the quantum dots away from their intended locations. We now show that CODA can be used to determine an optimal device scale.
	
	We again use the semiclassical Thomas-Fermi approximation \cite{Stopa:1996p13767} and the WKB approximation \cite{shankar1994principles} to model the devices. Since it is relatively difficult to determine the gate voltages needed to achieve single-electron occupancies in each dot, we choose a starting point for our simulations with 30 electrons in all three dots, and tunnel couplings that yield transmission coefficients of 0.01 between the left and middle dots and the middle and right dots. In each device, we then use the CODA protocol to automatically tune gate voltages to achieve single-electron occupation of each dot, while keeping the transmission coefficients constant, exploiting CODA's ability to automatically tune voltages. Additionally, we require the orbital energy splitting of each dot to be 1 meV or more, as consistent with recent experiments. Orbital energy splittings are calculated using a 2D finite-difference Schr{\"o}dinger solver; see Appendix B for details.
	
	The minimized voltage $L_1$ norms required for the auto-tuning process in each device are plotted in Fig~\ref{fig3}(b). Comparing these results, we see that the voltage $L_1$ norm is minimized for the device labeled (d), with a paddle width of approximately 80 nm, suggesting that this device is optimal from a control standpoint.
	
	The voltage changes for the auto-tuning protocol and the resulting electron charge density distributions are shown in Figs.~\ref{fig3}(c)-\ref{fig3}(e). For the optimal device, shown in Fig.~\ref{fig3}(d), the number of gates with voltage changes is relatively small, indicating good voltage sparsity. For devices smaller than the optimal device, voltage sparsity is still maintained, as a benefit of small device size. However, voltage moderation is not, with large voltage changes required on multiple electrodes, as shown in Fig.~\ref{fig3}(c). 
	For devices larger than optimal, voltage moderation is maintained, but the solution is no longer sparse, as shown in Fig.~\ref{fig3}(e). The constraints on dot occupation, dot energy and transmission coefficients lead to constraints on the size and relative position of the dots. In smaller devices, such as those shown in Fig.~\ref{fig3}(c),(d), the dots can be formed underneath the paddle electrodes and still meet these requirements. In larger devices, such as the device in Fig.~\ref{fig3}(e), to achieve the required quantities of interest, the dots can no longer form under the paddle gate electrodes, with the right-most dot forming under the nominal right tunnel-barrier electrode as indicated by the arrows. This misalignment between electrodes and dots, which has also been observed in experimental devices \cite{Veldhorst:2015aa,Ward:2016aa,Watson:2018aa}, leads to solutions that are less sparse than in the smaller devices.
	
	\section{Conclusion}
	We have introduced a protocol for the Compressed Optimization of Device Architectures, 
	which determines the optimal voltage changes for a given device operation by minimizing their $L_1$ norm.
	We have demonstrated the effectiveness of this scheme by considering its application to semiconductor nanoelectronic quantum dot systems.
	As devices continue to grow in complexity, such automated control schemes will be essential for design and operation.
	Our protocol is computationally efficient to implement, 
	and it provides a systematic approach for achieving local and sparse control.
	Through realistic semiclassical simulations of {multi}-dot devices, we have illustrated how the CODA scheme can be used for quantitative benchmarking and 
	device development. {While the current work focuses on quantum dot geometries, we note that the CODA protocol could also be applied to other device geometries, including donor-bound qubits, using simulation tools appropriate for those systems.}
	This method provides a path toward the rational design and operation of extensible quantum nanodevices.
	
	\section*{Acknowledgments}
	The authors thank C.~King, J.~B.~Aidun, J.~Moussa, N.~T.~Jacobson, R.~P.~Muller, and C.~Tahan for useful comments and discussions. This work was supported in part by ARO (W911NF-12-1-0607, W911NF-17-1-0274), NSF (PHY-1104660), and the Vannevar Bush Faculty Fellowship program sponsored by the Basic Research Office of the Assistant Secretary of Defense for Research and Engineering and funded by the Office of Naval Research through Grant No. N00014-15-1-0029. The views and conclusions contained in this document are those of the authors and should not be interpreted as representing the official policies, either expressed or implied, of the Army Research Office (ARO), or the U.S. Government. The U.S. Government is authorized to reproduce and distribute reprints for Government purposes notwithstanding any copyright notation herein.  This paper describes objective technical results and analysis. Sandia National Laboratories is a multimission laboratory managed and operated by National Technology \& Engineering Solutions of Sandia, LLC, a wholly owned subsidiary of Honeywell International Inc., for the U.S. Department of Energy’s National Nuclear Security Administration under contract DE-NA0003525.  The authors gratefully acknowledge support from the Sandia National Laboratories Truman Fellowship Program, which is funded by the Laboratory Directed Research and Development (LDRD) Program. This paper describes objective technical results and analysis. Any subjective views or opinions that might be expressed in the paper do not necessarily represent the views of the U.S. Department of Energy or the United States Government.
	
	\section{Appendix}

	\subsection{CODA protocol}
	Here, we provide further details about the CODA protocol described in the main text. 
	The simulated device is considered to be a nonlinear function $\hat S: \mathcal V \rightarrow \mathcal Q$, where $\mathcal V$ is the space of electrode voltages and $\mathcal Q$ is the space of quantities of interest (\emph{e.g.}, dot occupations, dot energies, transmission coefficients).
	Suppose that we have $n$ quantites of interest and $m$ electrode voltages, and that $m>n$, so the system is underconstrained. 
	We first identify a starting point of experimental interest $(\mathbf v_\text{op}^0, \mathbf q_\text{op}^0)$ such that $\hat S (\mathbf v_\text{op}^0)=\mathbf q_\text{op}^0$, and a target quantity of interest $\mathbf q_\text{target}$. It is our goal to find the vector $\delta \mathbf v_\text{tot}$ with the minimum $L_1$ norm which satisfies $\hat S (\mathbf v_\text{op}^0 + \delta \mathbf v_\text{tot})=\mathbf q_\text{target}$.
	
	The following are the steps taken at the $(i+1)^\text{th}$ iteration of CODA.
	\begin{enumerate}
		
		\item Given a working point $(\mathbf v_\text{op}^i, \mathbf q_\text{op}^i)$, consider a set of linearly independent, small voltage variations 
		$\{\boldsymbol \epsilon_1,\boldsymbol \epsilon_2,...,\boldsymbol \epsilon_m\}$ ($\boldsymbol{\epsilon}_j \in \mathcal{V}$) about the current working point. In the simulations described in the main text, we assumed took $\boldsymbol \epsilon_j$ to correspond to a voltage change of 0.1 mV on the $j^\text{th}$ electrode.
		Perform $m$ simulations ${\hat S} (\mathbf v_\text{op}^i + \boldsymbol{\epsilon}_j)$ to obtain
		the resulting $m$ $\mathbf q_\text{op}^i  + \delta \mathbf q_j$.
		From the collection of $\{\boldsymbol \epsilon_1,\boldsymbol \epsilon_2,...,\boldsymbol \epsilon_m\}$ and the associated $\{\delta \mathbf q_1,\delta \mathbf q_2,...,\delta \mathbf q_m\}$, construct the Jacobian matrix $\mathbf J_i$ using the method of least squares. For small $\delta \mathbf v$,  $\hat S (\mathbf v_\text{op}^i + \delta \mathbf v) \approx \mathbf q_\text{op}^i + \mathbf J_i \cdot \delta \mathbf v$.
		
		\item 
		Using a convex program (such as the matrix-free conic optimization implemented in the CVXPY package \cite{JMLR:v17:15-408}), find the $\delta \mathbf v_i$ that minimizes $ \left| \left| \mathbf v_\text{op}^i - \mathbf v_\text{op}^0 + \delta \mathbf v_i \right| \right|_1$ subject to the constraint  $\mathbf q_\text{target} = \mathbf q_\text{op}^i + \mathbf J_i \cdot \delta \mathbf v_1$. The vector $\mathbf v_\text{op}^i - \mathbf v_\text{op}^0 + \delta \mathbf v_i$ is the total change in voltage from the initial working point ($\mathbf v_\text{op}^0$).
		
		\item The voltage change vector $\delta \mathbf v_i$ defines a search direction, similar to the gradient used in nonlinear gradient descent. Evaluate ${\hat S} (\mathbf v_\text{op}^i + \delta \mathbf v_i)$; if the quantities of interest move closer to the target, \emph{i.e.}, $ \left| \left| {\hat S} (\mathbf v_\text{op}^i + \delta \mathbf v_i) - \mathbf q_\text{target} \right| \right|_2 < \left| \left| \mathbf q_\text{op}^i - \mathbf q_\text{target} \right| \right|_2$, then define $\mathbf v_\text{op}^{i+1}  = \mathbf v_\text{op}^i + \delta \mathbf v_i$. If not, then replace $\delta \mathbf v_i$ with $\delta \mathbf v_i/2$, and repeat this step. Continue until the quantities of interest move closer to the target.
	\end{enumerate}
	We repeat this process until $\left| \left| \mathbf q_\text{op}^i - \mathbf q_\text{target} \right| \right|_2$ is below a specified threshold value. For the simulations described in the main text, we assumed a threshold value of 0.01. For a detailed explanation of the units of this vector, see the following section.
	
	To decrease the total number of device simulations in CODA, one can replace step 1 in the protocol described above by Broyden's method \cite{Broyden-1965}. This method finds an approximate Jacobian matrix $\mathbf J_{i+1}$ by combining the Jacobian obtained in the previous iteration $\mathbf J_{i}$ and the nonlinear error from the previous step: $\hat S (\mathbf v_\text{op}^i + \delta \mathbf v) - (\mathbf q_\text{op}^i + \mathbf J_i \cdot \delta \mathbf v)$. Although this eliminates the need for explicit Jacobian formation, using Broyden's method often does increase the number of iterations required for convergence. In many cases, using this method leads to an overall speedup. However, in cases where sparse control cannot be achieved, we find that the number of iterations required for convergence increases dramatically, which negates any potential speed-up.
	
	{As with all "hill-climbing'' nonlinear optimization algorithms, there is no guarantee that the local optimum found by the CODA protocol is a global optimum. 
		However, because CODA is a regularized optimization protocol, the solution with the globally minimal $L_1$ norm is by definition "close" in control space to the starting point, and therefore it is likely that the solution found by CODA is the global minimum.
		While it is certainly possible to devise systems in which the CODA protocol does not find the global minimum, all indications are that the solutions found in the main text are indeed global minima.
		In principle, one could better ensure global optimization for these systems by implementing a version of CODA which uses a basin-hopping protocol to sample across several local minima.}

	\subsection{Simulation details}
	We perform semi-classical Thomas-Fermi calculations \cite{Stopa:1996p13767} using the COMSOL Multiphysics software package to solve a nonlinear Poisson equation in three dimensions.
	We use zero-field boundary conditions on all sides of the simulated domain, with the exception of the bottom of the SiGe buffer, which is grounded.
	We assume the following heterostructure profile for all the modeled devices. This profile is consistent with the accumulation-mode devices described in Refs. \onlinecite{Wu:2014p11938} and \onlinecite{Kawakami:2014p666}:
	200~nm of Si$_{0.7}$Ge$_{0.3}$ (with dielectric constant $\varepsilon=13.19$),
	a 10~nm Si quantum well ($\varepsilon=11.7$),
	30~nm of Si$_{0.7}$Ge$_{0.3}$,
	10~nm of Al$_2$O$_3$ ($\varepsilon=9.0$),
	a 10~nm layer of metallic electrodes embedded in the dielectric,
	80~nm of Al$_2$O$_3$,
	and a second 10~nm layer of metallic electrodes, followed by vacuum.
	Midway within the Si quantum well, we define a plane of charge with the charge density given by
	\begin{equation}
	\sigma_{\text{2D}}(x,y) = -2 \times 2 \times \frac{e m_{\textrm{eff}}(U(x,y)+E_F)}{2\pi \hbar^2} \times \theta(U(x,y)+E_F),
	\end{equation}
	where $e$ is the charge of an electron, $m_{\textrm{eff}} = 0.19\, m_{\text{electron}}$ is the transverse effective mass of a conduction electron in silicon, $U(x,y)$ is the strength of the electrostatic potential energy as a function of position, $E_F$ is the Fermi energy (which we take to be at ground), and $\theta(x)$ is the step function. The two prefactors account for the spin and valley degeneracies.
	
	The dot occupations are calculated via integrating the charge density found with the Thomas-Fermi approximation. Transmission coefficients between dots are calculated by finding the center of charge of each dot, and applying the WKB approximation \cite{shankar1994principles}
	across a straight line connecting the two charge centers. Orbital dot energies are calculated via a 2D finite-difference Schr{\"o}dinger solver in the plane of charge, using the electrostatic confinement potential obtained from the Thomas-Fermi analysis, and the transverse effective mass of a conduction electron in silicon.
	
	Our CODA protocol requires all the components of the voltage vector to have the same units (and comparable magnitudes, for numerical stability).
	The quantities of interest considered in our simulations were electron occupations and tunnel barrier heights.
	When populating our vectors in the space $\mathcal Q$, we use the units of electron number for dot occupation, meV for dot energy and we take the logarithm of the transmission coefficient, divided by 1000, since tunnel couplings can vary by orders of magnitude as a function of gate voltage. These units were chosen to ensure rapid convergence.
	
	The details of the initial working point used in the analysis of the 8-dot device are given in the supplemental file \verb|8DotDevice.txt|. 
	In this file, the physical attributes are listed first.
	The dot occupations are expressed in electron numbers, and the transmission coefficients are unitless. Voltages are given for each electrode, with the following labeling convention defined with respect to Fig.~1(g) of the main text.
	Beginning with the upper layer of electrodes, Electrode~1 is in the lower-right corner of the schematic, and the ordering proceeds clockwise.
	In the lower layer of gates, Electrode~5 is in the lower-right corner, and the ordering again proceeds clockwise.
	
	Similar details for working points on the devices shown in Fig.~3 are given in the supplemental file \verb|TripleDot.txt|. The dot occupations are given in numbers of electrons, the transmission coefficients are scaled as before, and the dot orbital energies are given in meV. Here, the labeling convention for the electrodes begins with the upper layer at the electrode in the upper right corner of the schematic and proceeds clockwise.
	On the lower layer of electrodes, the labeling begins at the electrode in the upper-left corner and proceeds clockwise.
	
	\subsection{Tunnel rates and transmission coefficients}
	Following Ref. \onlinecite{landauLifshitz} one can derive that the tunnel coupling $\Delta$ between two one-dimensional simple harmonic oscillators with frequencies $\omega_1$ and $\omega_2$ is approximately
	\begin{equation}
	\Delta \approx \frac{1}{2\pi}T_{WKB} \hbar \sqrt{\omega_1\omega_2},
	\end{equation}
	where $T_{WKB}$ is the transmission coefficient calculated via the WKB approximation. For quantum dots with orbital energies of $\sim$1 meV, a transmission coefficient of 0.01 corresponds with a tunnel coupling of $\Delta \approx 1.6$ $\mu$eV, or a tunnel rate of $\Delta/h \approx 400$ MHz.

\end{document}